\documentstyle[prl,twocolumn,aps,psfig]{revtex}

\begin{document}
\draft
\preprint{}

\title{Isotropic Conductivity of Two-Dimensional \\
Three-Component Symmetric Composites}

\author{Leonid G. Fel , Vladimir Sh. Machavariani and  David J. Bergman}
\address{School of Physics and Astronomy,\\ Raymond and Beverly Sackler
Faculty of Exact
Sciences\\Tel Aviv University, Tel Aviv 69978, Israel\\
}

\date{\today}

\maketitle

\def\be{\begin{equation}}
\def\ee{\end{equation}}
\def\p{\prime}

\subsection*{Abstract}
\begin{abstract}

The effective dc-conductivity problem of isotropic, two-dimensional
(2D), three-component,
symmetric, regular composites is considered. A simple cubic equation with
one free
parameter for $\sigma_{e}( \sigma_1,\sigma_2,\sigma_3)$ is suggested
whose solutions automatically have all the exactly known properties of
that function.
Numerical calculations on four different symmetric, isotropic,
2D, three-component, regular structures show a non-universal behavior
of
$\sigma_{e}(\sigma_1,\sigma_2,\sigma_3)$ with an essential dependence
on
micro-structural details, in contrast with the analogous two-component
problem. The applicability of the cubic equation to these
structures
is discussed. An extension of that equation to the description of
other types of 2D three-component structures is suggested,
including the case of random structures.

Pacs: 72.15.Eb, 72.80.Tm, 61.50.Ah

\end{abstract}
\noindent
\narrowtext
\section{Introduction}
\label{introduction}

The \nolinebreak classical duality transformation of two-di\-men\-sio\-nal
(2D) heterogeneous composites,
discovered by Keller \cite{keller64} and independently by Dykhne
\cite{dykhne70}, has been
applied in a restricted set of physical contexts. The dual symmetry is
based upon the observation that any 2D divergence-free field, when
rotated locally at each
point by $90^{o}$, becomes curl-free, and vice versa.  This leads to
the result that static
effective physical properties of 2D infinite heterogeneous composites,
like electric conductivity
${\widehat \sigma_{e}}$, thermal conductivity, dielectric permittivity,
as well as some other
static properties, satisfy some exact relationships which follow from
the similarity between dual
problems. For example, in the case of a two-component composite, {\it a
universal square-root law
behavior} of the bulk effective transport characteristics was
demonstrated \cite{dykhne70},\cite{mendel75}.
It is also known\footnote{It seems to be strange but we have not found
throughout the papers concerned
with this subject any published proof of this statement. Such proof is
so useful that we give it
in Appendix.} that this transformation leads to exact results in 2D
infinite composites made of an
arbitrary number of components. Dykhne \cite{dykhne70} gave sufficient
conditions that, when satisfied
by the component conductivities $\sigma_i$, lead to exact results for
the isotropic bulk effective
conductivity $\sigma_e(\sigma_i)$.
We know of only one attempt \cite{emets98} to consider a 3-component 2D
composite with a doubly
periodic arrangement of two kinds of circular inclusions embedded into
the matrix. The effective permittivity
was obtained
using the dipole approximation for the inclusion polarizations.
But we do not know any
{\it rigorous} results obtained for
any 2D composite microstructure, made of 3 (or more) components, with
arbitrary component conductivities.
Apparently, this is not an accident but reflects the disappearance of
commutativity
in symmetric groups when we upgrade from the $S_2$ permutation group to
the $S_n$ permutation group.

In this paper we consider the effective isotropic conductivity problem
for $3$-component
2D infinite composites with translational order, with a microstructure
that is {\em symmetric} in the 3
components. We formulate an approach based on the {\it conjecture} of
the {\it algebraicity} of
$\sigma_{e}(\sigma_1,\sigma_2,\sigma_3)$ and its general  properties.
We have found that the  algebraic
equation of minimal order where all these properties
can be satisfied is a {\it cubic}  equation, which contains 1 free
parameter,  with coefficients
made of the independent invariants of $S_3$. This equation is in
agreement with Dykhne' result (see
Eq.\ (\ref{dyk1})). Its predictions are compared with numerical
solutions for $\sigma_e$ in some regular 3-component microstructures.
It is also extended so as to apply to other types of 3-component
composites, including random microstructures.


\section{TWO-DIMENSIONAL THREE-COLOR COMPOSITES}
\label{part2}

The effective dc-conductivity problem for $n$-component symmetric, 2D,
infinite
composites with
translational order, can be reformulated with the help of {\it n-color}
plane groups.

Color groups are generalizations of the classical crystallographic
groups. Different
colors may correspond, for example, to different chemical species or,
more generally, to different
values of a physical property which is defined as a tensor of $k$-rank
: scalar - density $\rho$ ,
vector - magnetic spin {\bf m} , tensor of 2-nd rank - conductivity
${\widehat \sigma}_{ij}$,
tensor of 3-rank - piezoelectric modulus ${\widehat d}_{ijn}$, etc.

Every $n$-color plane group has its origin in one
of the $N_1=$17 regular ({\it color-blind}\,) plane
groups. The number $N_n$ of {\it n-color} plane groups is a
non-monotonic function of $n$:
$N_2=46$ \cite{belov56}; $N_3=23$; $N_4=96$; $N_5=14$; $N_6=90$
\cite{senech79},
\cite{shwarz80}. The plane groups $n \leq 60$ are tabulated in
\cite{wiet82}.

Only 10 of the 23 3-{\it color} plane groups have a 3-fold rotation
axis:

\noindent
5 lattice equivalent
groups
 - ${\sf P3(L)|P1(L)}$, ${\sf P6(L)|P2(L)}$, ${\sf P31m(L)|Cm(L)}$,
 ${\sf P3m1(L)|Cm(L)}$,
${\sf P6mm(L)|Cmm(L)}$ and 5 class equivalent groups  - ${\sf
P3(L)|P3(L^{'})}$,
${\sf P6(L)|P6(L^{'})}$, ${\sf P31m(L)|P3m1(L^{'})}$,${\sf
P3m1(L)|P31m(L^{'})}$,
${\sf P6mm(L)|P6mm(L^{'})}$ where ${\sf L^{'}}$ is a possible
sub-lattice\footnote{We follow the notations of Ref. \cite{shwarz80} for
three-color plane groups ${\sf G_{col}}=
{\sf G(L)|G^{'}(L^{'})}$ which means that {\sf G} is the geometrical,
or Fedorov, plane
group and the subgroup ${\sf G^{'}} \subset {\sf G}$ of index 3
contains operations that keep the
first color fixed. We do not specify here the relationship between the
lattice {\sf L} and its
sub-lattice ${\sf L^{'}}$.} of ${\sf L}$ invariant under rotations of
order 3 \cite{shwarz80}. All
these groups are compatible only  with hexagonal Bravais lattices.
The 3-fold rotational symmetry makes the effective conductivity in
structures governed by
those groups isotropic. This follows from the Hermann theorem
\cite{herm34} about $k$-rank
tensors in media with an inner symmetry which includes a
rotation axis of
highest order $r,\;r>k$. Despite their different geometries, all these
structures have one
important property in common: They are all invariant under the full
permutation group $S_3$
which  exchanges the colors, therefore they are related to the
permutational
crystallographic color groups \cite{birman82}.

\section{EFFECTIVE CONDUCTIVITY $\sigma_{\lowercase{e}}
(\sigma_1,\sigma_2,\sigma_3)$ and ITS  ALGEBRAIC PROPERTIES}
\label{part3}

A direct way to solve the dc-conductivity problem for an $n$-component
composite begins
with the local field equations
\begin{equation}
\nabla\times\;{\bf E}({\bf r})=0\;,\;\;\;\nabla\cdot\;{\bf J}({\bf
r})=0\;,\;\;
{\bf J}({\bf r})=\sigma ({\bf r})\;{\bf E}({\bf r})\;,
\label{rot}
\end{equation}
along with appropriate boundary conditions for the electrical
potential. The local
conductivity $\sigma ({\bf r})$ is a discontinuous function
$\sigma ({\bf r})=\sigma_i$, if ${\bf r} \in \Delta_i\;,\;i=1,2,3$
where $\Delta_i$ is a
homogeneous part of the composite with constant conductivity
$\sigma_i$.
The isotropic effective conductivity $\sigma_{e}$ can be defined via
Ohm's law for the
current ${\bf J}_e$ and the field ${\bf E}_e$  averaged over the system
\begin{equation}
{\bf J}_e=\sigma_{e}\;{\bf E}_e\;,\;{\bf J}_e=\frac{1}{S}
\int{\bf J}({\bf r}) d S\;,\;
{\bf E}_e=\frac{1}{S} \int{\bf E}({\bf r}) d S\;.
\label{om1}
\end{equation}
Except for a medium with trivial 1D inhomogeneities (a layered medium)
an exact solution
of this problem does not exist for any regular or random structure. At
the same time the
function $\sigma_{e}(\sigma_i)$ must have the following general
properties which we are
going to exploit:

${\sf 1.\;Homogeneity\; of\; 1-st\; order}$
\begin{equation}
\sigma_{e}( k\; \sigma_1,\;k\; \sigma_2,\;k\; \sigma_3)=
k\;\sigma_{e}( \sigma_1,\;\sigma_2,\;\sigma_3)\;.
\label{prop1}
\end{equation}
This follows from the linearity of the static Maxwell equations
(\ref{rot}) and from the
definitions of the average current and field (\ref{om1}).

${\sf 2.\;Permutation\; invariance}$
\begin{equation}
\sigma_{e}({\widehat {\cal
P}_l}\;\{\sigma_1,\;\sigma_2,\;\sigma_3\})=\;
\sigma_{e}(\sigma_1,\;\sigma_2,\;\sigma_3)\;.
\label{prop2}
\end{equation}
where ${\widehat {\cal P}_l},\;l=1,...,6$ is a permutation operator of
the indices
$\{1,2,3\}$ (or of 3 colors). The six operators ${\widehat {\cal P}_l}$
form the
non-commutative group $S_3$. The existence of permutation invariance
presumes that the 3
components are distributed with equal volume fractions
$p_1=p_2=p_3=1/3$.

${\sf 3.\;Duality}$
\begin{equation}
\sigma_{e}(\sigma_1,\;\sigma_2,\;\sigma_3) \times
\sigma_{e}(\frac{1}{\sigma_1},\;\frac{1}{\sigma_2},\;\frac{1}{\sigma_3})=1
\;.
\label{prop3}
\end{equation}
See Eq. (\ref{dual4}).

${\sf 4.\; Compatibility}$
\begin{equation}
\sigma_{e}(\sigma,\;\sigma,\;\sigma)=\sigma \;.
\label{prop4}
\end{equation}
The last formula does not follow from the previous ones but reflects a
natural requirement.

Dykhne \cite{dykhne70} proved a theorem for a symmetric composite of 3
components,
namely
\begin{equation}
\sigma_{e}(\sigma_1,\;\sigma_2,\;\sigma_3)=\sqrt{\sigma_1
\sigma_2}\;,\;\;\;\mbox{if}\;\;\;
\sigma_3^2=\sigma_1 \sigma_2\;.
\label{dyk1}
\end{equation}
The last formula represents the only known rigorous result for
3-{\it color} 2D isotropic composites.
It can easily be shown to follow from (\ref{prop2}) and (\ref{prop3}).
Let us mention one more conclusion which follows from (\ref{dyk1})
\begin{equation}
\sigma_{e}(\sigma,\;0,\;0)=0\;,
\label{dyk2}
\end{equation}
which reflects the percolation property of such a composite. In fact,
it is easy to
construct 3-{\it color} 2D isotropic composites with the more
specialized  property
\begin{equation}
\sigma_{e}(\sigma_1,\;\sigma_2,\;0)=0\;.
\label{dyk3}
\end{equation}
Fig.\ref{f1234}d illustrates such a structure with {\it traps}, i.e. a
network of closed,
simply connected loops, which enclose plaquettes of the two other
colors. If any color
denotes an insulator, then the effective conductivity of the composite
is clearly zero.
The 2-{\it color} 2D square checkerboard has a similar property. It
differs from the
case of Fig.\ref{f1234}d by the value of the percolation threshold. The
previous example
(\ref{dyk3}) shows {\sf a non-universal behavior} of
$\sigma_{e}(\sigma_i)$ even for
symmetric isotropic microstructures, with essential dependence on
micro-structural
details, in contrast to the 2-{\it color} 2D symmetric
isotropic composites,
where $\sigma_e=\sqrt{\sigma_1\sigma_2}$ always \cite{dykhne70}.

We will look for $\sigma_{e}(\sigma_1,\sigma_2,\sigma_3)$ satisfying
the requirements
(\ref{prop1} - \ref{prop4}) among algebraic functions.
This choice is inspired by the fact that the symmetric 2-{\it color} 2D
isotropic
composite generates a quadratic equation for
$\sigma_{e}(\sigma_1,\sigma_2)$.
Another motivation is the situation which exists vis-a-vis some
discrete 2D models in
statistical mechanics, where a duality transformation exists that is
similar to the one
invoked here. Such a transformation exists for the Ising and Potts
models
\cite{baxt82}, where the critical point equations
are algebraic --- quadratic equation for the Ising model on a self-dual
square lattice and
cubic equations for the Potts model on the mutually dual triangle and
honeycomb lattices.

According to the ``fundamental theorem of symmetric functions''
\cite{olver98} the symmetric
group $S_3$ has 3 algebraically independent homogeneous
invariants (basic
invariants)
\begin{eqnarray}
I_1&=&\sigma_1+\sigma_2+\sigma_3\;,\;
I_2=\sigma_1 \sigma_2 + \sigma_2 \sigma_3 + \sigma_3 \sigma_1\;,\;\nonumber \\
I_3&=&\sigma_1 \sigma_2 \sigma_3 \;,
\label{inv1}
\end{eqnarray}
which satisfy the  obvious restrictions
\[ \frac{1}{3} I_1\;\geq\; \frac{I_2}{I_1}\;\geq
\;3\;\frac{I_3}{I_2}\;\;,\;\;\;
\frac{1}{3} I_1\;\geq\; \sqrt[3]{I_3}\;\geq\; 3\;\frac{I_3}{I_2}\;,\]
and can be used as independent variables instead of
$\sigma_1,\sigma_2,\sigma_3$.
The difference $I_2-I_1\;\sqrt[3]{I_3}$ can have either sign.
The function $\sigma_{e}(I_1,I_2,I_3)$ now satisfies (\ref{prop2})
automatically.

It can be shown that the algebraic equation of minimal order for the
function $\sigma_{e}(I_1,I_2,I_3)$,
which satisfies all basic requirements, is a {\it cubic} equation of
the form
\begin{equation}
\sigma_{e}^3+A\;I_1\; \sigma_{e}^2-A\; I_2\; \sigma_{e}-I_3=0\;,
\label{eqv1}
\end{equation}
where $A$ is a free parameter responsible for the non-universality,
and $\sigma_{e}$ is a value bounded from both above and below \cite{wien12}
\begin{equation}
\frac{1}{3}I_1 \geq \sigma_{e} \geq 3\;\frac{I_3}{I_2}\;.
\label{rest1}
\end{equation}
It is easy to show that the equation (\ref{eqv1}) satisfies all basic
requirements
(\ref{prop1}-\ref{prop4}) and automatically satisfies Dykhne's theorem
(\ref{dyk1})
independently of $A$. Indeed, the requirements (\ref{prop1},
\ref{prop2}, \ref{prop4}) one can
check immediately. The duality property (\ref{prop3})
follows when we notice that (\ref{eqv1}) is equivalent to the
cubic equation for
$\frac{1}{\sigma_e}(\frac{1}{\sigma_1},\;\frac{1}
{\sigma_2},\;\frac{1}{\sigma_3})$:
\begin{eqnarray}
\frac{1}{\sigma_e^3} +
&A&\; (\frac{1}{\sigma_1}+\frac{1}{\sigma_2}+\frac{1}{\sigma_3})\cdot
\frac{1}{\sigma_e^2} -\nonumber \\
&A&\; (\frac{1}{\sigma_1 \sigma_2}+\frac{1}{\sigma_2 \sigma_3} +
\frac{1}{\sigma_3 \sigma_1})\cdot \frac{1}{\sigma_e} -
\frac{1}{\sigma_1 \sigma_2 \sigma_3}=0 \;.\nonumber
\label{invers}
\end{eqnarray}
The property (\ref{dyk1}) can be proven by straightforward
substitution $\sigma_3=\sqrt{\sigma_1 \sigma_2}$ into (\ref{eqv1}) .

As we will see later, $A$ reflects not only the plane group of a color
tessellation but also the shape of the elementary cell. To illuminate
the meaning of $A$
let us put into (\ref{eqv1}) $A=\frac{1}{3}$. After simple
algebra one obtains the following equation
\begin{equation}
\frac{\sigma_{e}-\sigma_1}{\sigma_{e}+\sigma_1}+
\frac{\sigma_{e}-\sigma_2}{\sigma_{e}+\sigma_2}+
\frac{\sigma_{e}-\sigma_3}{\sigma_{e}+\sigma_3}=0\;,
\label{brug}
\end{equation}
which coincides
with the Bruggeman effective medium approximation for a symmetric,
3-component, 2D composite \cite{brug35}.

From restrictions (\ref{rest1}) and after straightforward
manipulations, one can show that $A$ is bounded
from below
\begin{equation}
\infty \;\geq\;A\;\geq\;0\;.
\label{boun2}
\end{equation}
It is noteworthy that the lower bound ensures that (\ref{eqv1}) has only
one positive root, thus avoiding the possibility of multiple
physical solutions.
More accurate estimation of the lower bound for $A$, using the
Hashin-Shtrikman \cite{hash62}
exact bounds $\sigma_{HS}^{\pm}$ for isotropic conductivity
$\sigma_{e}$ of 3-component 2D composite, does not change
(\ref{boun2}).
Indeed, for $\sigma_1 > \sigma_2 > \sigma_3$ one has
\begin{equation}
\frac{1}{3}I_1 \geq \sigma_{HS}^{+} \geq
\sigma_{e} \geq \sigma_{HS}^{-} \geq 3\;\frac{I_3}{I_2}\;,
\label{rest2}
\end{equation}
where
\begin{eqnarray}
\sigma_{HS}^{+}&=&\sigma_1\cdot(1+\frac{4(\sigma_2\sigma_3-\sigma_1^2)}
{\sigma_2\sigma_3+3\sigma_1(\sigma_2+\sigma_3)+5\sigma_1^2})\;,\nonumber \\
\sigma_{HS}^{-}&=&\sigma_3\cdot(1+\frac{4(\sigma_1\sigma_2-\sigma_3^2)}
{\sigma_1\sigma_2+3\sigma_3(\sigma_1+\sigma_2)+5\sigma_3^2})\;.\nonumber
\label{hash11}
\end{eqnarray}
Due to the ambiguity of the general case, we consider the special case
$\sigma_1 > \sigma_2=\sigma_3$.
Here we have
\[ \sigma_{HS}^{+}=\sigma_1\cdot\frac{5+r}{1+5r}\;,\\
\sigma_{HS}^{-}=\sigma_3\cdot\frac{1+2r}{2+r}\;,\;\;r=\frac{\sigma_1}{\sigma_3}
\geq 1\;. \]
We can then derive the following lower bounds from (\ref{eqv1}),
(\ref{rest1})
\begin{eqnarray}
&&A>-\frac{1+5r}{1+2r}\cdot\frac{1+r}{(2+r)^2}\;,\nonumber \\
&&A>-\frac{r}{1+5r}\cdot\frac{2+r}{1+r}\cdot
\frac{1+18r+r^2}{1+16r+r^2}\;.\nonumber
\label{bound}
\end{eqnarray}
Since the last inequalities must fit all cases including $r\rightarrow
\infty$, we return to (\ref{boun2}).

\section{EFFECTIVE CONDUCTIVITY of REGULAR STRUCTURES:
NUMERICAL RESULTS}
\label{part4}

\begin{figure}[h]
\vspace{-1 cm}
\psfig{figure=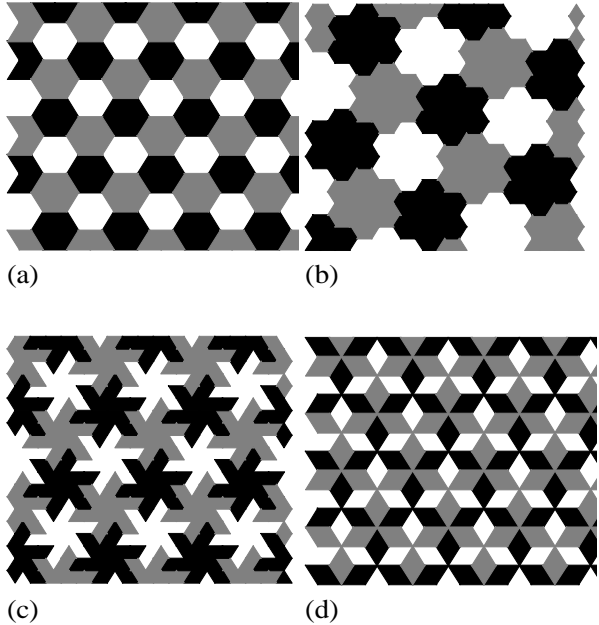,height=12cm,width=9cm}
\vspace{-2 cm}
\caption{3-{\it color} 2D composites:
a) {\sf He} $-{\sf G_{col}}={\sf P6mm(L)|P6mm(L^{'})}$,
b) {\sf Fl} $-{\sf G_{col}}={\sf P6(L)|P6(L^{'})}$,
c) {\sf Co} $-{\sf G_{col}}={\sf P6(L)|P6(L^{'})}$,
d) {\sf Rh} $-{\sf G_{col}}={\sf P6mm(L)|P6mm(L^{'})}$.}
\label{f1234}
\end{figure}

In the present section we will study numerically four different
infinite 2D 3-{\it color} class equivalent
regular structures of ${\sf P6mm(L)|P6mm(L^{'})}$ and ${\sf
P6(L)|P6(L^{'})}$ types. In order to avoid cumbersome
notations for these structures, we will use the following symbols:
{\sf He} (honeycomb) (Fig.\ref{f1234}a) ; {\sf
Fl} (flower) (Fig.\ref{f1234}b) ; {\sf Co} (cogrose) (Fig.\ref{f1234}c)
; {\sf Rh} (rhombus) (Fig.\ref{f1234}d).

\begin{figure}[t]
\vspace{0.5 cm}
\centerline{
\psfig{figure=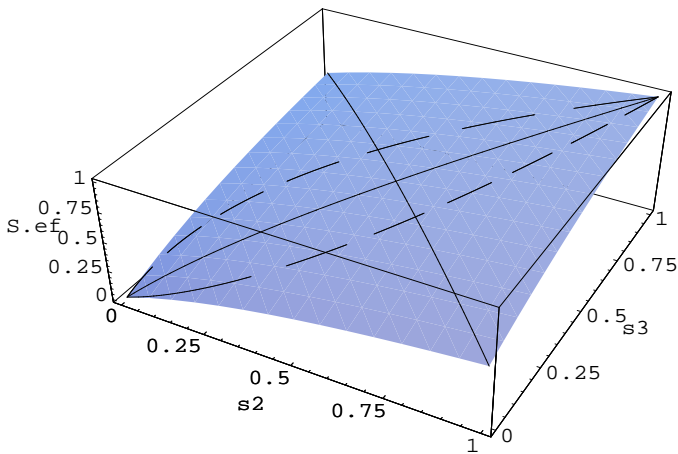,height=6cm,width=9cm}
}
\vspace{0.5 cm}
\caption{Typical shape of the surface
$\sigma_{e}(\sigma_2^{\prime},\sigma_3^{\prime})$ for
3-{\it color} 2D isotropic composites.
The loop where Dykhne's theorem is satisfied is shown by the {\it
dashed} lines.}
\label{hex3d}
\end{figure}

Due to the homogeneity (\ref{prop1}) of Eq.\ (\ref{eqv1}) one
can
rescale $\sigma_{e}(\sigma_1,\sigma_2,\sigma_3)$ defining a new
function
$\sigma_{e}^{\prime}(\sigma_2^{\prime},\sigma_3^{\prime})$ where
$\sigma_i^{\prime}=
\sigma_i/\sigma_1,\;i=e,2,3$. A typical shape of the surface defined by
$\sigma_{e}^{\prime}(\sigma_2^{\prime},\sigma_3^{\prime})$ is shown in
Fig. \ref{hex3d}.
This surface always contains the following points
\[ \sigma_{e}^{\prime}(0,0)=0\;,\;\;\sigma_{e}^{\prime}(1,1)=1\;,\;\;
\sigma_{e}^{\prime}(\sigma,\sqrt{\sigma})=\sqrt{\sigma}\;.\]
The last equality is related to the Dykhne theorem and
generates a loop on the surface where (\ref{dyk1}) is satisfied.
This loop is common for all such surfaces,
i.e. they intersect each other on it.

We computed $\sigma_{e}^{\prime}(\sigma_2^{\prime},\sigma_3^{\prime})$
for the structures mentioned above:
{\sf He, Fl, Co, Rh}, and then extracted the parameter $A$
corresponding to those structures.

Before discussing these results we describe briefly the numerical
algorithm which was used.
This algorithm deals with hexagonal Bravais lattices represented as a
grid of equilateral triangles.
A subdivision procedure produces a set of $N\times N$ small similar triangles,
the center
of each one of them connected with the centers of 3 nearest
neighboring cells by simple resistors.
The resistor connecting cells $i$ and $j$ has a resistance
equal to
$(\sigma_i+\sigma_j)/(2\sqrt{3}\sigma_i \sigma_j)$, where $\sigma_i$ is
the conductivity
of the cell $i$. Translational symmetry of the composite is
reflected by imposing
periodic boundary conditions for the currents. The algorithm solves up
to 10$^6$ linear
equations arising for the subdivided 3-{\it color} elementary cell. Our
computational
procedure always
gives a sequence of effective conductivities $\sigma_{N}^{-}$,
which converges monotonically (as $N \rightarrow$ 1000)
from below. This fact was established by treating the solvable
case following from
Dykhne's theorem, where $\sigma_e$ is known exactly. To get a sequence of
upper bounds $\sigma_{N}^{+}$ we then used
the {\it duality} property, i.e. simulating the dual problem with
conductivities
$1/\sigma_2^{\prime},\;1/\sigma_3^{\prime}$ and consequently obtaining
a monotonic convergence
from above for the sequence $\sigma_{N}^{+}$.

\begin{figure}[h]
\vspace{-1 cm}
\centerline{
\psfig{figure=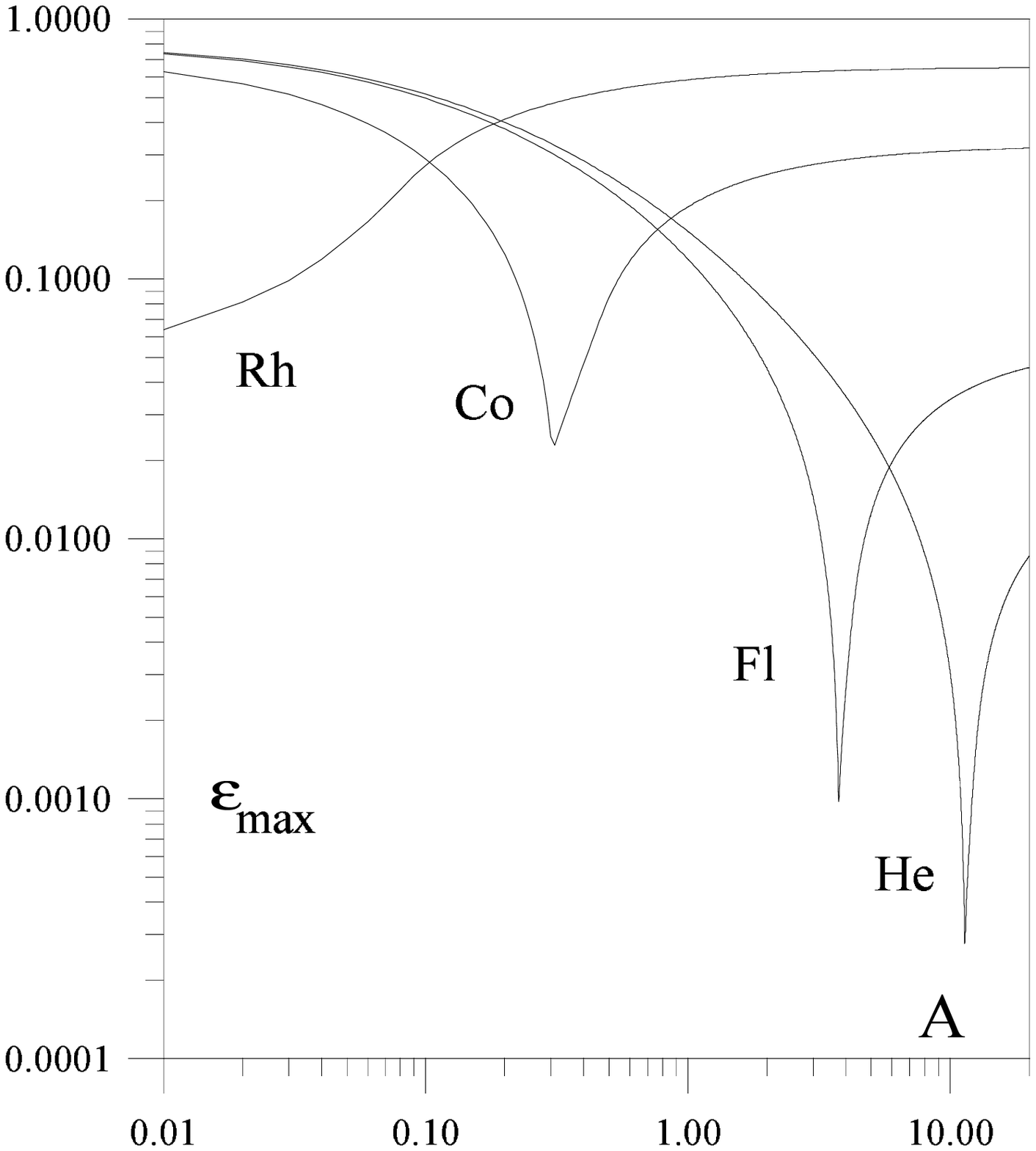,height=10cm,width=10cm}
}
\vspace{-1 cm}
\caption{The fitting curves $\epsilon_{max}(A)$ for 4 different
structures.
The best value of $A$ is determined by the minimum in each curve. Those
values are:
{\sf He} - $A_{He}=11.37,\; \epsilon_{max}^{He}$=.00026 ;
{\sf Fl} - $A_{Fl}=3.76, \; \epsilon_{max}^{Fl}$=.001 ;
{\sf Co} - $A_{Co}=.305,\; \epsilon_{max}^{Co}$=.022 ;
{\sf Rh} - $A_{Rh}=0, \;\epsilon_{max}^{Rh}$=.065.}
\label{all}
\end{figure}

The simulations were done on the 4 characteristic curves
\footnote{Because of the divergence of the calculational procedure at
$\sigma_i^{\prime}=0$, the first of the
characteristic curves was calculated using $\sigma_2^{\prime}=0.001$,
for which the results can still be trusted.}
 on the surface where it intersects with the planes:
\begin{equation}
1)\; \sigma_2^{\prime}=0.001\; ,\;\; 2)\; \sigma_2^{\prime}=1\; ,\;\;
3)\;
\sigma_2^{\prime}=\sigma_3^{\prime}\; ,\;\;
4) \;\sigma_2^{\prime}+\sigma_3^{\prime}=1\;.
\label{curve1}
\end{equation}
These curves reflect the behavior of the surface relatively well in
accordance with the
chosen type of composite. From the assumption of the algebraic
nature (\ref{eqv1}) of
those curves we were able to extract, for every type of composite, a
corresponding
parameter $A$ by the following procedure.
In a wide range of $A$, for each calculated point
($\sigma_i^{\prime},\;i=2,3$) on the
plane, relative deviations $\epsilon$ were calculated
\begin{equation}
\epsilon(A,\sigma_i^{\prime})=\left\{ \begin{array}[c]{ccl}
\frac{\sigma_e(A,\sigma_i^{\prime})-\sigma_{N}^{+}}
{\sigma_e(A,\sigma_i^{\prime})}
&
{\rm if} &
\sigma_{N}^{+}<\sigma_e(A,\sigma_i^{\prime}) \\
 0 & {\rm if} &
 \sigma_{N}^{-}<\sigma_e(A,\sigma_i^{\prime})<\sigma_{N}^{+} \\
\frac{\sigma_{N}^{-}-\sigma_e(A,\sigma_i^{\prime})}
{\sigma_e(A,\sigma_i^{\prime})}
&
{\rm if} &
\sigma_e(A,\sigma_i^{\prime})<\sigma_{N}^{-} \nonumber
\end{array}\right. \nonumber
\end{equation}
and a maximal value of those deviations $\epsilon_{max}(A)$ was
determined for every value of $A$ by
scanning over the entire area 0$<\sigma_{2,3}^{\prime}<$1 .
We then determined the best value of $A$ by minimizing the function
$\epsilon_{max}(A)$. As shown in
Fig. \ref{all}, $A$ for the {\sf He} and {\sf Fl} structures is
determined by very sharp minima.
The other two minima are not as sharp.

As one could expect, the {\sf Rh} structure has $A=0$ since this is the
unique value of
$A$ for which the solution of equation (\ref{eqv1}) has the features
typical of the structures with traps. The values of $A$ which minimize
$\epsilon_{max}(A)$ for the
other structures are listed in the caption of Fig. \ref{all}.

\begin{figure}[t]
\psfig{figure=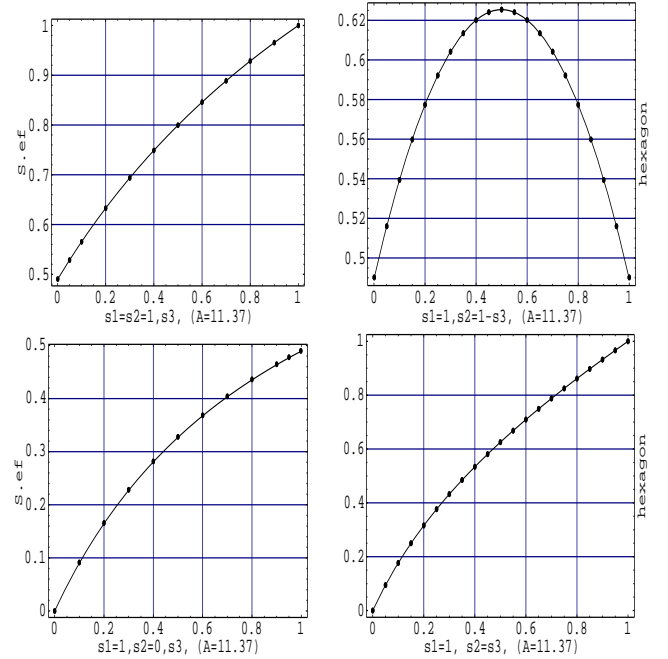,height=9.5cm,width=9cm}
\caption{Numerical results
$\sigma_{N}^{+}(\sigma_2^{\prime},\sigma_3^{\prime}),\;
\sigma_{N}^{-}(\sigma_2^{\prime},\sigma_3^{\prime})$ ({\it black dots})
and analytic
solution  $\sigma_{e}^{\prime}(\sigma_2^{\prime},\sigma_3^{\prime})$ of
cubic
equation (\ref{eqv1}) with $A$=11.37 ({\it bold line}) at the four
sections (\ref{curve1}) of
the surface corresponding to the {\bf He} - structure.}
\label{hexg3}
\end{figure}
\begin{figure}[h]
\psfig{figure=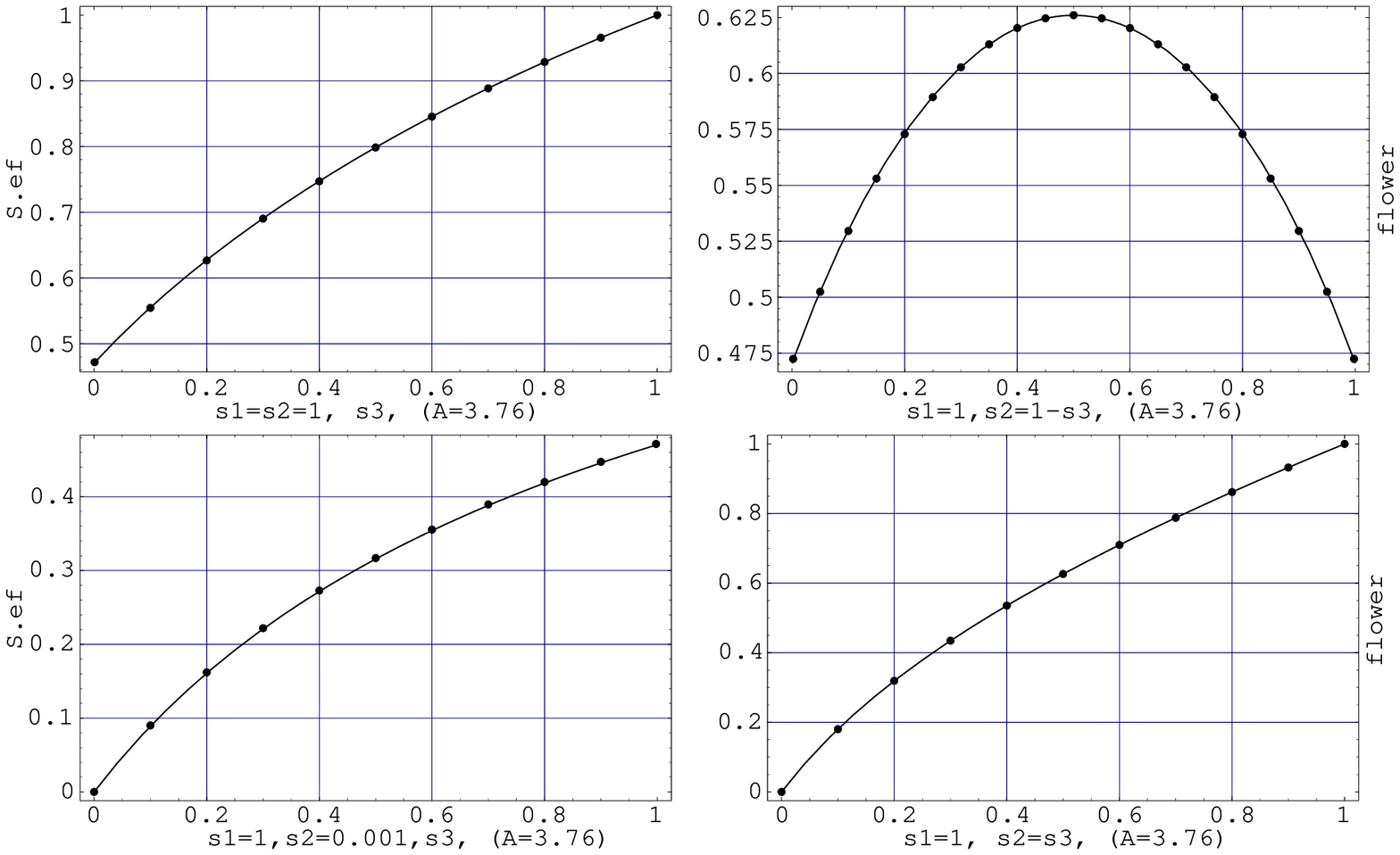,height=9.5cm,width=9cm}
\caption{Numerical results
$\sigma_{N}^{+}(\sigma_2^{\prime},\sigma_3^{\prime}),\;
\sigma_{N}^{-}(\sigma_2^{\prime},\sigma_3^{\prime})$
({\it black dots})  and analytic solution
$\sigma_{e}^{\prime}(\sigma_2^{\prime},\sigma_3^{\prime})$
of cubic equation (\ref{eqv1}) with $A$=3.76 ({\it bold line}) at the
four sections (\ref{curve1}) of the
surface corresponding to the {\bf Fl} - structure.}
\label{rosg3}
\end{figure}
\begin{figure}[h]
\psfig{figure=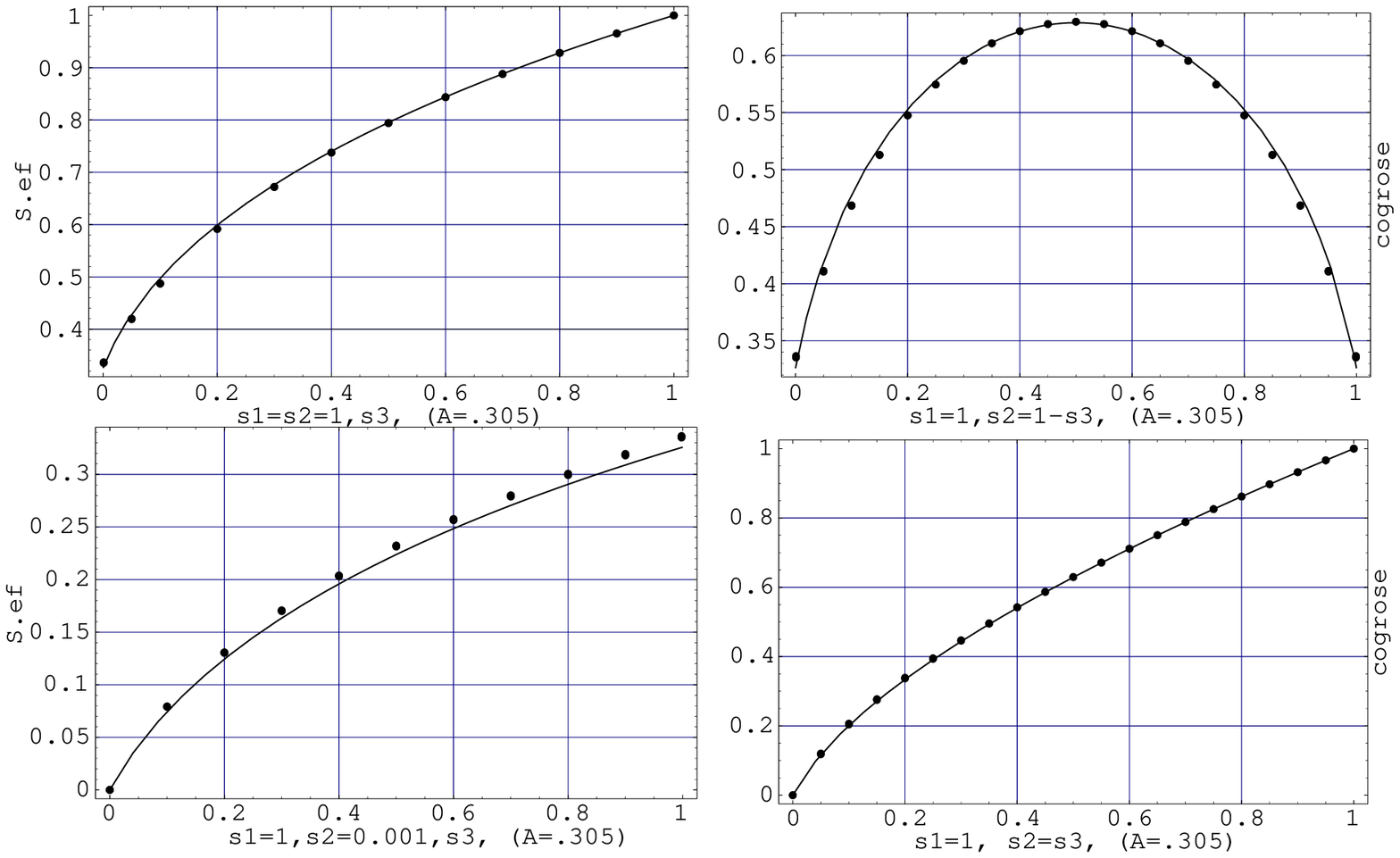,height=9.5cm,width=9cm}
\caption{Numerical results
$\sigma_{N}^{+}(\sigma_2^{\prime},\sigma_3^{\prime}),\;
\sigma_{N}^{-}(\sigma_2^{\prime},\sigma_3^{\prime})$ ({\it black dots})
and analytic solution
$\sigma_{e}^{\prime}(\sigma_2^{\prime},\sigma_3^{\prime})$ of cubic
equation (\ref{eqv1}) with
$A$=0.305 ({\it bold line}) at the four sections (\ref{curve1}) of the
surface corresponding to the {\bf Co} - structure.}
\label{maltg3}
\end{figure}
\begin{figure}[h]
\psfig{figure=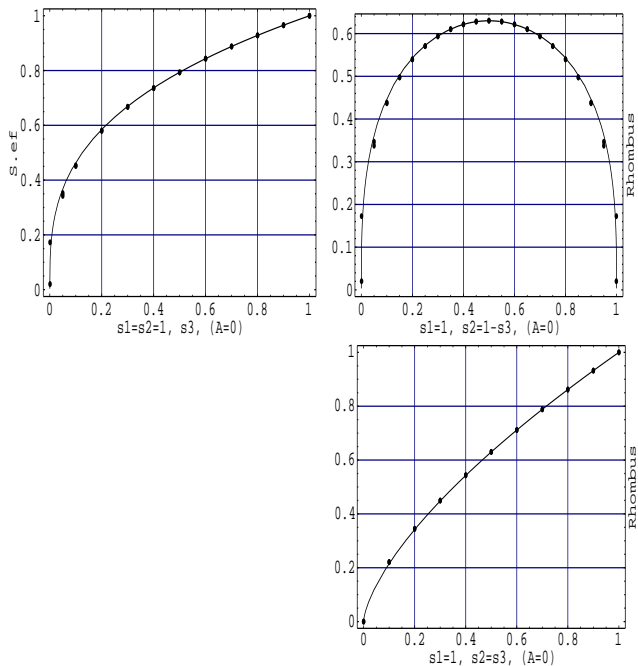,height=9.5cm,width=9cm}
\caption{Numerical results
$\sigma_{N}^{+}(\sigma_2^{\prime},\sigma_3^{\prime}),\;
\sigma_{N}^{-}(\sigma_2^{\prime},\sigma_3^{\prime})$ ({\it black dots})
and analytic solution
$\sigma_{e}^{\prime}(\sigma_2^{\prime},\sigma_3^{\prime})$ of cubic
equation (\ref{eqv1}) with
$A$=0 ({\it bold line}) at the three sections (\ref{curve1}) of the
surface corresponding
to the {\bf Rh} - structure. The fourth dependence
$\sigma_{e}^{\prime}(0,\sigma_3^{\prime})=$0 is omitted because of its
triviality.}
\label{rombg3}
\end{figure}

Figs. \ref{hexg3}--\ref{rombg3} show the computed results for upper
and lower bounds on the relative
bulk effective conductivity $\sigma_{e}^{\prime}$, i.e.
$\sigma_{N}^{+}(\sigma_2^{\prime},\sigma_3^{\prime})$
and $\sigma_{N}^{-}(\sigma_2^{\prime},\sigma_3^{\prime})$
at maximal $N$, for the 4
microstructures of Fig. \ref{f1234}.
Note that these upper and lower bounds 
often appear to coincide due to
insufficient resolution in the figures.
The axes in the figures are labeled with {\sf S.ef} for the relative
bulk effective conductivity
$\sigma_{e}^{\prime}$ and {\sf s1, s2, s3} for
the component relative
conductivities
$\sigma_1^{\prime}=1,\sigma_2^{\prime},\sigma_3^{\prime}$
respectively.

The pairs of almost merged points shown in Fig.\ \ref{rombg3}
correspond to barely
separated upper and lower bounds.


\noindent

\section{EFFECTIVE CONDUCTIVITY OF RANDOM STRUCTURES:
EXTENSION OF THE ALGEBRAICITY CONJECTURE}
\label{part5}

In this chapter we discuss briefly a possible extension of the
algebraicity conjecture for 2D three-component structures,
which are non-regular but macroscopically homogeneous.
On length scales large compared to the
inhomogeneities, we can characterize the macroscopic response
of such a medium by a single number, the plane effective
conductivity $\sigma_e$. A reexamination of the basic
properties (\ref{prop1}-\ref{prop4}) of $\sigma_{e}
(\sigma_1,\sigma_2,\sigma_3)$ shows
that one of them (\ref{prop2}) must be discarded,
while the other three (homogeneity of 1-st order,
duality, and compatibility) continue to be valid. It is important
to note that these properties hold irrespective of whether the
microstructure of the (isotropic) composite is ordered or disordered.

\begin{figure}[h]
\centerline{
\psfig{figure=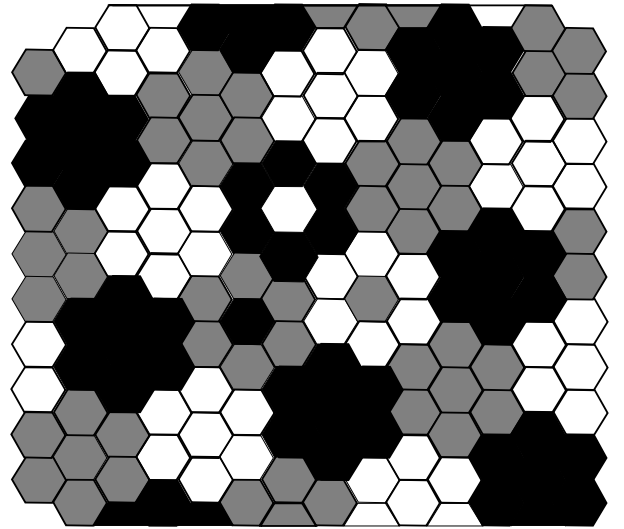,height=7cm,width=8cm}
}
\vspace{0.5 cm}
\caption{3-{\it color} 2D composite of {\sf Fl} structure
locally disturbed in the centers of only three plaquettes with
different colors.
Such color tessellation preserves an equal fraction of every color
and is invariant under the cyclic permutation group $C_3$,
but not under the full $S_3$ permutation group.
}
\label{circ3}
\end{figure}

Instead of Eq.\ (\ref{prop2}), which implies full symmetry under
the group $S_3$ of all 3-color
permutations, we first consider the case where
the structure is only symmetric
under the group $C_3$ of cyclic 3-color permutations:
1 $\rightarrow$ 2 $\rightarrow$ 3 $\rightarrow$ 1. An example of such
a microstructure is shown in Fig.\ \ref{circ3}. (Note the
arrangement of ``flowers'' near the center of that figure.)
It follows that
$\sigma_e(\sigma_1,\sigma_2,\sigma_3)$ is a cyclic permutation invariant
function. The group $C_3$ is a subgroup
of index 2 of the full permutation group $S_3$, which
characterizes all the
regular structures {\sf He}, {\sf Fl}, {\sf Co}, {\sf Rh} discussed in the
previous sections.

Being non-reflecting, the group $C_3$ has more basic invariants
than the group $S_3$ [see Eq.\ (\ref{inv1})], namely \cite{olver98}
\begin{eqnarray}
I_1(C_3)&=&\sigma_1+\sigma_2+\sigma_3\;,\;\;
I_2(C_3)=\sigma_1 \sigma_2 +\sigma_2 \sigma_3 +\sigma_3 \sigma_1
\;,\nonumber \\
I_3(C_3)&=&\sigma_1 \sigma_2 \sigma_3\;,\;
I_4(C_3)=(\sigma_1 - \sigma_2)(\sigma_2 - \sigma_3)(\sigma_3 -
\sigma_1)\;.\nonumber
\label{inv3}
\end{eqnarray}
Nevertheless, the additional
cubic invariant $I_4(C_3)$ cannot be incorporated into
a cubic equation for $\sigma_e$, because that would violate the
duality requirement. Thus, we are lead back to
the cubic equation (\ref{eqv1}), which
is dictated not only by the strong requirement (\ref{prop2}) of the
full permutation invariance $S_3$, but even
by the milder requirement of cyclic permutation invariance $C_3$. 

In the case of a random composite, we usually characterize
the microstructure by a statistical distribution function of
the local conductivity, which can be either $\sigma_1$,
$\sigma_2$, or $\sigma_3$ at any point. Such a description
results in an ensemble of representative structures, each one
with its own form for the function $\sigma_e(\sigma_1,\sigma_2,\sigma_3)$.
If we assume that the distribution function has the permutation symmetry of
either $S_3$ or $C_3$, then the ensemble average of
$\sigma_e(\sigma_1,\sigma_2,\sigma_3)$ will have that symmetry too,
even though individual samples may violate it. It follows that,
the same considerations which led us to stipulate the form
(\ref{eqv1}), for the minimal polynomial equation that $\sigma_e$
could satisfy for regular structures, also lead to that same
equation for random structures, if the statistical model
for those structures is invariant under either $S_3$ or $C_3$.
Numerical tests of this conjecture remain to be performed.

\section{CONCLUSION}
\label{part6}

In the present paper we have introduced the algebraicity conjecture for
the effective 
conductivity problem of isotropic 2D three-component regular composites.
This
conjecture is based on the general properties which are satisfied by
the effective
conductivity $\sigma_e(\sigma_1,\sigma_2,\sigma_3)$. The algebraic
equation of minimal
order for $\sigma_e$ is a cubic equation with 1 positive free parameter
$A$ responsible
for the non-universality. This equation satisfies Dykhne's theorem
(\ref{dyk1})
independently of $A$ and has only one positive root,
thus avoiding the possibility of multiple physical solutions.
The value $A=1/3$ corresponds exactly to the Bruggeman effective medium
approximation for a 2D composite with 3 equally partitioned components.

We have found support for this conjecture
by numerical calculations on four different
infinite 2D 3-{\it color} class equivalent regular structures of ${\sf
P6mm(L)|P6mm(L^{'})}$
and ${\sf P6(L)|P6(L^{'})}$ types: {\sf He}, {\sf Fl}, {\sf Co}, {\sf
Rh}.
We have established that $\sigma_{e}(\sigma_1,\sigma_2,\sigma_3)$ is a
non-universal
function with essential dependence on the microstructure even for
totally symmetric structures:

1) The cubic equation (\ref{eqv1}) with $A$=11.37 governs the
conductivity problem in
{\sf He} structure with a very high precision $\epsilon_{max}\simeq
10^{-4}$.

2) There is good agreement ($\epsilon_{max}\simeq 10^{-3}$) between
the cubic equation
with $A$=3.76 and numerical results for the {\sf Fl} structure.

3) In the {\sf Co} structure the estimated value $A$=.305 is near $1/3$,
which would
follow from the Bruggeman effective medium approximation. This may
indicate some similarity
between the conducting properties of a 3-component random
microstructure and those of the
ordered {\sf Co} structure.

4) The {\sf Rh} structure needs special attention. It belongs to the
family of
structures with {\it unicolor} traps, i.e. with structures where
the presence of just one non-conducting
($\sigma_1=0$) component is enough to make the composite an insulator
($\sigma_e=0$).
In percolation theory this corresponds to a threshold $p_3=1/3$, in
contrast with a
2-{\it color} composite where $p_2=1/2$. If the cubic equation is
valid for this structure with $A$=0, then $\sigma_{e}=\sqrt[3]{\sigma_1
\sigma_2 \sigma_3}$.
Unfortunately, these computations are unable to resolve this question
for the {\sf Rh}
microstructure.

5) The different microstructures which belong to a common plane
symmetry group
({\sf He}, {\sf Rh} - ${\sf P6mm(L)|P6mm(L^{'})}$, {\sf Fl}, {\sf Co} -
${\sf P6(L)|P6(L^{'})}$) are characterized by distinct values of
$A$. This means that
this parameter has a topological nature
and is sensitive to more than just the symmetry properties of the
elementary cell or the plane group of the entire color lattice.

Finally we discussed a possible extension of the algebraicity
conjecture to other types of 2D three-component structures. 
First we showed that even if the microstructure
is only invariant under the $C_3$ subgroup of the $S_3$
permutation group, a cubic equation for $\sigma_e$ must
still have the form (\ref{eqv1}).
We then extended this conjecture also to the case of random
microstructures, described by any statistical model that is
invariant under either $S_3$ or $C_3$.
To test this idea, it would be useful to
do numerical calculations on such structures.
This is left for future investigations.

\acknowledgements

We would like to thank J.L. Birman, A.M. Dykhne, I.M. Khalatnikov,
Y.B. Levinson and A. Voronel for helpful discussion.

This research was supported in part by grants from the Tel Aviv
University Research
Authority, from the Gileadi Fellowship program of the Ministry of
Absorption
of the State of Israel (LGF), and from the Aaron Gutwirth Foundation,
Allied Invest.\ Ltd.\ (VSM).


\appendix
\renewcommand{\theequation}{\thesection\arabic{equation}}
\section{}
\label{appendix1}
\setcounter{equation}{0}

We now give a simple proof of the duality relations for a 2D
isotropic medium composed of an arbitrary number $n$ of isotropic
components.

Suppose a 2D medium with a continuous distribution of conductivity
$\sigma({\bf r})$ is subjected to an average electric field ${\bf
E}_e$. The system of equations consists of
Ohm' law
\begin{equation}
{\bf J}({\bf r})={\widehat \sigma}({\bf r})\;{\bf E}({\bf r})\;,
\label{Ohm}
\end{equation}
and the equations for local fields
\begin{equation}
\nabla\times\;{\bf E}({\bf r})=0\;,\;\;\;\nabla\cdot\;{\bf J}({\bf
r})=0\;
\label{rot_div}
\end{equation}
with appropriate boundary conditions on the electrical potential.

We are interested in the relation between the current ${\bf J}_e$
averaged over the system,
${\bf J}_e=S^{-1} \int{\bf J}({\bf r}) d S$ and the averaged field
${\bf E}_e=S^{-1}
\int{\bf E}({\bf r}) d S$. By virtue of the linearity of (\ref{Ohm},
\ref{rot_div}) this
relation will also be linear
\begin{equation}
{\bf J}_e={\widehat \sigma_{e}}\;{\bf E}_e\;,\;\;\;{\widehat
\sigma}_e={\widehat
\sigma}_e\;\{{\widehat \sigma}({\bf r})\}\;.
\label{Ohmef}
\end{equation}
where the tensor of the bulk effective conductivity
${\widehat \sigma}_e={\widehat \sigma}_e\;\{{\widehat \sigma}({\bf
r})\}$ is actually a tensorial
functional. In the case of $n$ homogeneous anisotropic components, the
latter becomes a tensorial
function ${\widehat \sigma}_e\;({\widehat \sigma}_1,{\widehat
\sigma}_2,...,{\widehat \sigma}_n)$.
Further simplification arises when all components are isotropic ---
${\widehat \sigma}_e\;(\sigma_1,\sigma_2,...,\sigma_n)$ and, finally
when the entire composite is also
an isotropic medium --- $\sigma_e\;(\sigma_1,\sigma_2,...,\sigma_n)$.

In order to transform to the  dual problem
we rotate the $x,y\;$-components of ${\bf J}$ and ${\bf E}$ by $90^o$
in the plane
\begin{eqnarray}
&&{\bf J}^{\prime}({\bf r})={\widehat R}\;{\bf E}({\bf r})\;,\;\;
{\bf E}^{\prime}({\bf r})={\widehat R}\;{\bf J}({\bf r})\;,\;\;
{\bf J}^{\prime}({\bf r})={\widehat \sigma}^{\prime}({\bf r})\;{\bf
E}^{\prime}({\bf r})\;,\nonumber \\ 
&&{\widehat \sigma}^{\prime}({\bf r})={\widehat R}\;{\widehat
\sigma}^{-1}({\bf r})\;
{\widehat R}^{-1}\;,\;\;{\widehat R}=
\left(\begin{array}{cc}
0 & -1\\
1 & 0\end{array}\right).
\label{trans1}
\end{eqnarray}
Eqs. \ref{rot_div} are thereby transformed as follows
\[ \nabla\cdot\; {\bf J}^{\prime}({\bf r})=\nabla\times\; {\bf E}({\bf
r})=0\;,\;\;
\nabla\times\; {\bf E}^{\prime}({\bf r})=\nabla\cdot\; {\bf J}({\bf
r})=0\;,\]
while ${\bf J}_e^{\prime}$ and ${\bf E}_e^{\prime}$ are connected by
\begin{equation}
{\bf J}^{\prime}_e={\widehat \sigma}^{\prime}_e\;{\bf
E}^{\prime}_e\;,\;\;
{\widehat \sigma}^{\prime}_e={\widehat R}\;{\widehat
\sigma}_e^{-1}{\widehat R}^{-1}\;,
\label{trans2}
\end{equation}
where the bulk effective conductivity tensor of the dual problem
${\widehat \sigma}^{\prime}_e$
is defined in accordance with (\ref{Ohmef})
\begin{equation}
{\widehat \sigma}^{\prime}_e={\widehat \sigma}_e \{{\widehat
\sigma}^{\prime}({\bf r})\}=
{\widehat \sigma}_e\;\{{\widehat R}\;{\widehat \sigma}^{-1}({\bf
r}){\widehat R}^{-1}\}\;.
\label{trans3}
\end{equation}
The last two equations (\ref{trans2},\ref{trans3}) give a duality
relation
\begin{equation}
{\widehat \sigma}_e\;\{{\widehat R}\;{\widehat \sigma}^{-1}({\bf
r}){\widehat R}^{-1}\}\cdot
{\widehat R}\cdot{\widehat \sigma}_e\;\{{\widehat \sigma}({\bf
r})\}\cdot{\widehat R}^{-1}=
{\widehat I}\;,
\label{dual1}
\end{equation}
where ${\widehat I}$ is the unit matrix. If, instead of the continuous
fields $\sigma({\bf r})$ we
deal with {\sf anisotropic media} composed of $n$ {\sf homogeneous
anisotropic components}, then
\begin{eqnarray}
{\widehat \sigma}_e\;
({\widehat R}\;{\widehat \sigma}^{-1}_1\;{\widehat R}^{-1},{\widehat
R}\;{\widehat \sigma}^{-1}_2\;
{\widehat R}^{-1},...,
{\widehat R}\;{\widehat \sigma}^{-1}_n\;{\widehat R}^{-1}) \cdot
{\widehat R} \cdot \nonumber \\
\cdot {\widehat \sigma}_e\;({\widehat \sigma}_1,{\widehat
\sigma}_2,...,
{\widehat \sigma}_n) \cdot {\widehat R}^{-1}={\widehat I}\;.
\label{dual11}
\end{eqnarray}
In the case of an {\sf anisotropic composite} with $n$ {\sf homogeneous
isotropic components}
we will have
\begin{equation}
{\widehat
\sigma}_e\;(\sigma^{-1}_1,\sigma^{-1}_2,...,\sigma^{-1}_n)\cdot
{\widehat R}\cdot{\widehat
\sigma}_e\;(\sigma_1,\sigma_2,...,\sigma_n)\cdot{\widehat R}^{-1}=
{\widehat I}\;.
\label{dual2}
\end{equation}
The principal values of ${\widehat \sigma}_e$ satisfy Keller's theorem
\begin{eqnarray}
{\widehat
\sigma}_e^{xx}(\sigma^{-1}_1,\sigma^{-1}_2,...,\sigma^{-1}_n)\cdot
{\widehat
\sigma}_e^{yy}(\sigma_1,\sigma_2,...,\sigma_n)=1\;,\\ \nonumber
{\widehat
\sigma}_e^{yy}(\sigma^{-1}_1,\sigma^{-1}_2,...,\sigma^{-1}_n)\cdot
{\widehat \sigma}_e^{xx}(\sigma_1,\sigma_2,...,\sigma_n)=1\;.
\label{dual3}
\end{eqnarray}
In general, the directions of the principal axes depend on the values
of $\sigma_1,\sigma_2,$ etc.
But when the symmetry of the microstructure is sufficiently high, those
directions will be fixed by
that symmetry. Finally, in the case of an {\sf isotropic composite},
the last equations reduce to the {\it
self-duality} relation
\begin{equation}
\sigma_e(\sigma^{-1}_1,\sigma^{-1}_2,...,\sigma^{-1}_n)\cdot
\sigma_e(\sigma_1,\sigma_2,...,\sigma_n)=1\;.
\label{dual4}
\end{equation}

\end{document}